\documentclass[prb,aps,twocolumn,amsmath,amssymb,floatfix,
superscriptaddress]{revtex4}

\usepackage[dvips]{graphics}

\usepackage{color}

\date{\today}

\begin{document}

\title{Non-volatile reconfigurable spin logic device: Parallel operations}

\author{Moumita Patra}

\email{moumita.patra19@gmail.com}

\affiliation{Department of Physics, Indian Institute of Technology Bombay,
Mumbai, Maharashtra-400 076, India}

\author{Alok Shukla}

\email{shukla@phy.iitb.ac.in}

\affiliation{Department of Physics, Indian Institute of Technology Bombay,
Mumbai, Maharashtra-400 076, India}

\author{Santanu K. Maiti}

\email{santanu.maiti@isical.ac.in}

\affiliation{Physics and Applied Mathematics Unit, Indian Statistical
Institute, 203 Barrackpore Trunk Road, Kolkata-700 108, India}

\begin{abstract}

A new proposal is given for designing a non-volatile, completely spin logic device, that can be reprogrammed for different functional 
classical logical operations. We use the concept of bias driven spin dependent circular current and current induced magnetic field in 
a quantum ring under asymmetric ring-to-electrode interface configuration to implement all the Boolean operations. We extend our idea 
to build two kinds of parallel computing architectures for getting parallelized operations, all at a particular time. For one case, 
different kinds of parallel operations are performed in a single device, whereas in the other type all the possible inputs of a logic 
gate are processed in parallel and all the outputs are read simultaneously. The performance and reliability are 
investigated in terms of power, delay and power-delay-product and finally the system temperature. We find that both the individual
and simultaneous logic operations studied here are much superior compared to the operations performed in different conventional logic
families like complementary metal oxide semiconductor (CMOS) logic, transistor-transistor logic (TTL), etc. The key advantage is that 
we can perform several logic operations, as many as we wish, repeating the same or different logic gates using a single setup, which 
indeed reduces wiring in the circuits and hence consumes much less power. Our analysis can be utilized to design optimized logic 
circuits at nano-scale level.

\end{abstract}

\maketitle

\section{Introduction}

New generation spin logic gates use up and down spin configurations of electrons as input and output states. The main advantage of 
a spin logic gate is that it can additionally store output, whereas in the conventional microprocessors, information need to be 
transferred to memory to prevent it from getting lost, as the electrically processed informations are volatile~\cite{cite1} in nature. 
Along with it, we can also expect several other key advantages like significantly lower power consumption, fast processing, higher 
integration densities together with non-volatility, and to name a few. These features are not involved with the 
conventional logic gates made up with semi-conducting materials~\cite{cite2,cite3,cite4,cite5} that belong to CMOS and TTL families.
The objectives of this work are to explore the possibilities of designing a non-volatile and reprogrammable spin logic gate with the help 
of a simple quantum ring, and to extend the idea in devising parallel logic gates.

The underlying physical mechanism of logical operations that we use here relies on the manipulation of bias driven circular current
and associated magnetic field in a quantum ring which is attached to external baths. At finite bias, a net circular current appears in
the ring conductor satisfying some conditions~\cite{cite6,cite7,cite8,cite9,cite10,cite11,cite12,cite13,cite14}, along with the transport 
current through the junction which is usually called as the drain current. Appearance of circular current in the conducting 
loop is directly connected to the effect of quantum interference of electronic waves passing through different arms of the junction. 
The circular current, on the other hand, induces a magnetic field at and near the center of the ring, and in our analysis we focus on 
the magnetic field that is developed at the center of the loop. Depending on applied bias and ring-to-electrode interface geometry, the 
induced magnetic field becomes reasonably large and in some specific situations, it may even reach to few milliTesla (mT) or even 
Tesla (T)~\cite{cite7,cite10,cite13}. The strength of the magnetic field is somewhat related with the loop area. 
Smaller loops yield much stronger magnetic field compared to the bigger ones for a fixed bias window. Existence of finite magnetic 
field triggers us to 
think about the logical operations by considering its effect on a (free) spin which is placed at the center of the loop. The induced 
magnetic field controls the orientation of this local (free) spin~\cite{cite7,cite13}. Exploiting this fact, here we design classical 
Boolean operations where all the inputs and outputs are spin based. Our device is made up of a magnetic quantum ring, with an embedded 
\begin{figure}[ht]
{\centering\resizebox*{8cm}{2.3cm}{\includegraphics{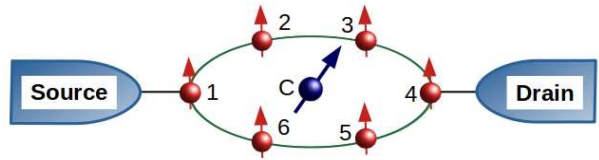}}\par}
\caption{(Color Online). Basic setup for the construction of logic gates where a magnetic nano ring is coupled to source and
drain in such a way that the upper and lower arms acquire equal lengths. Each site of the ring contains a net spin, oriented along
$Z$-direction, as shown by the red arrow. A free magnetic site is embedded at the center of the ring, whose net spin is aligned
in an arbitrary direction (blue arrow).}
\label{fig1}
\end{figure}
free spin at its center, where the ring is directly coupled to two semi-infinite non-magnetic source (S) and drain (D) electrodes 
(see Fig.~\ref{fig1}).

In the magnetic ring, all the lattice sites possess a net spin and among them the orientations (i.e., up or down) of one or two 
spins, depending on the one-input or two-input logic gates, are considered as the logic inputs. When an electron having a particular 
spin (up or down) gets incident from the source end, it interacts with the magnetic sites of the ring, which is known as spin-spin 
exchange interaction. Because of this, spin channels are separated, and we can get finite spin dependent junction current as well as 
circular current in the loop. That circular current induces a magnetic field which is utilized to adjust the orientation of the free 
spin. For the symmetrically connected ring i.e., when the status of the upper and lower arms are identical, circular current vanishes 
and as a consequence so does magnetic field. Under this situation the alignment of the local spin (not along the $Z$ direction) remains 
unchanged which we refer as $0$ output, while for the case of finite magnetic field, the spin is oriented along $+Z$ which is denoted 
as output $1$. As the output conditions are purely spin-dependent, information can also be stored without giving any extra effort which 
makes the output a non-volatile one. This is another important advantage over the charged based ones. In this communication, we 
successfully establish five logical operations (OR, NOT, XOR, XNOR, NAND), employing the above mentioned symmetry condition, and the 
key feature is that all these logical operations are performed by reprogramming a single setup.

Designing of logic gates where everything is spin based is of great impact, and the efficiency will be enhanced significantly if we can
achieve the parallel operations as well. Though some efforts have been made to fabricate nano-scale logic gates, very less amount of works is 
available so far where parallel operations are explored~\cite{datta1,cite16a,cite16b,cite16c,cite16d,mp1,mp2}, and thus further probing with 
a comprehensive discussion is undoubtedly required to have suitable functional operations. This is the primary motivation behind our work.
Here, we propose a general concept and specific implementation of individual and parallelized logic devices in two-terminal systems where 
all parts are spin based. Two kinds of parallel operations are given. In one case, the logical operations are performed in one setup. 
While, for the other case all the input conditions are implemented at a particular time and all the possible outputs are read simultaneously. 
Till now a single work is available, to the best of our concern, where Fresch {\em et al.} have `experimentally' demonstrated how to
achieve logic operations where all the inputs are processed in parallel and the outputs are read at the same time~\cite{cite16a}. In our
work, we extended the idea to design the same with non-volatile, spin-based programmability. To fulfill it, we take a logic gate (say, 
OR gate) as a unit, and add them in parallel where all the input spins align according to four input conditions i.e., (0,0), (0,1), (1,0), 
and (1,1) and all the corresponding outputs are read at a time. For the entire analysis, we define the input $0$ by up spin, while the 
input condition $1$ is mentioned by the down spin one. To have the input conditions $0$ and $1$, we need to align 
the spins in the magnetic ring selectively, and several prescriptions are available to control single electron spin. For instance, using 
radio frequency pulses the spin can be manipulated~\cite{sr1,sr2,sr3}, though in this case relatively larger time scale is required to 
operate the spin. On the other hand, the manipulation can be made much faster such as in the picosecond or femtosecond time scale with 
the help of optical pulses~\cite{sr4,sr5,sr6}. In another pioneering work Press {\em et al.} have shown that the selective tuning of 
electron spin is possible within the spin's coherence time by means of ultrafast laser pulses~\cite{sr7}. With the availability of these 
various sophisticated prescriptions, we strongly 
believe that the alignment of selective spins in our magnetic quantum ring can be adjusted to implement the desired logical operations.

Here it is relevant to note that the concept of bias driven circular current in a conducting loop in designing individual logic gates 
has already been reported in a very recent work~\cite{cite12}. A cyclic molecule is clamped between two contact electrodes to form a 
nanojunction where the input signals are given with the help of two external gate electrodes. Only the individual logic operations are 
explored, without providing any simultaneous logic operations, and all the logic functions are completely charge based. Moreover the 
concept of `spin dependent bias driven circular current' was not clear during the proposition of logic functions given in Ref.~\cite{cite12}.
Determination of spin dependent circular current involves the proper definition of spin current components in individual bonds connecting
the neighboring lattice sites. First it came into light with a detailed description in Ref.~\cite{cite14}, and in the present work we 
essentially use this concept of spin dependent bias driven circular current to implement individual as well as simultaneous logic operations. 
This is completely new and has not been attempted so far in literature. Another new concept is that, contrary to the earlier recent 
proposal of designing parallel logic gates~\cite{mp1} where at most two logic functions can be obtained simultaneously at the two outgoing
terminals, here we can perform a large number of logic operations, as many as we wish, repeating the identical or dissimilar logic gates, 
considering a single setup. The limitation of two logical operations in the earlier proposition was mainly due to the fact that the output
response is measured in terms of transport current, but this issues is completely eliminated in our present case where we measure the
output signal by bias driven spin dependent circular current. Because of this fact multiple operations are performed, all at a time, and 
the wiring in the circuits largely 
decreases which eventually reduces power consumption significantly. In Ref.~\cite{mp1}, the spin-dependent scattering is obtained by means
of spin-orbit (SO) coupling. Since this coupling is usually too weak, it is quite hard to produce a considerable separation between the up 
and down spin channels, and this is the primary requirement to have better spin dependent transport phenomena. This issue can be eliminated
by considering a magnetic system, like what is taken in our present work, where spin dependent scattering is too large as the spin-spin 
exchange coupling is usually of the order of eV~\cite{exch1}. We believe that the concept of using bias driven spin dependent circular 
current in a magnetic quantum loop will be the most suitable prescription for designing efficient logic operations.

Along with the designing mechanisms of different logic gates, we put emphasis on the efficiency of such gates under different input
conditions. Both the reliability and performance of logic devices are tested in terms of power, delay and power-delay-product (PDP), and
finally the system temperature. The PDP, the product of power dissipation and the delay, is the fundamental quantity that measures the 
efficiency of a device, and it is the goal of the optimized design to obtain its lowest possible value. From our detailed analysis we
find that, both 
the individual and simultaneous logic operations are much superior compared to the traditional TTL and CMOS logic families. All the 
above mentioned quantities associated with the device performance viz, power, delay and PDP become extremely small for our systems, 
and, the key reason of getting smaller values is that here all the logic operations are performed based on `spin states', circumventing 
the use of `charges' for ON/OFF states as considered in conventional cases. The logical operations are also much less sensitive to the 
system temperature, which is another key aspect of our propositions.

The rest of the work is arranged as follows. In Sec. II we present the basic magnetic ring setup which is being utilized to have different 
logic functions and the Hamiltonian of the conducting junction. A brief theoretical prescription is also given in this section. All the 
essential results are thoroughly presented and analyzed in Sec. III which include both single and the simultaneous logic operations. At 
the end of results and discussion, we provide a brief outline of designing magnetic nanorings, for the sake of completeness of our analysis. 
Finally in Sec. IV we summarize our essential findings.

\section{Quantum System and Theoretical prescription}

Let us start with the basic setup, schematically illustrated in Fig.~\ref{fig1}, where a magnetic nano ring is directly coupled to
non-magnetic electrodes S and D. The total Hamiltonian of this nanojunction can be written as a sum
\begin{equation}
H= H_C + H_S + H_D + H_T
\label{eq1}
\end{equation}
where $H_C$, $H_S$, $H_D$ and $H_T$ represent the sub-Hamiltonians for the channel (C) (viz, the ring conductor), source (S), drain (D),
and the tunnel coupling between the ring and the electrodes, respectively. To express these Hamiltonians we use a tight-binding (TB) 
prescription within a nearest-neighbor hopping (NNH) approximation, which always gives a simple level of description, particularly at 
nano-scale level. The forms of the sub-Hamiltonians look like
\begin{equation}
H_C=\sum_n\mbox{\boldmath$c_n^{\dagger}$} \left(\mbox{\boldmath$\epsilon_n$} - \mbox{\boldmath$h_n.\sigma$} \right)
\mbox{\boldmath$c_n$} + \sum_n \left(\mbox{\boldmath$c_{n+1}^{\dagger} t c_n$} + h.c. \right)
\label{eq2}
\end{equation}
\begin{equation}
H_M=\sum_n \mbox{\boldmath$c_n^{\dagger} \epsilon_0 c_n$} + \sum_n \left(\mbox{\boldmath$c_{n+1}^{\dagger} t_0 c_n$} + 
\mbox{\boldmath$c_n^{\dagger} t_0 c_{n+1}$} \right)
\label{eq3}
\end{equation}
where $M=S, D$.
The meaning of different terms are as follows. \mbox{\boldmath$\epsilon_0$}$=\mbox{diag}(\epsilon_0,\epsilon_0)$ and 
\mbox{\boldmath$t_0$}$=\mbox{diag}(t_0,t_0)$, where $\epsilon_0$ and $t_0$ represent the on-site energy and NNH integrals in the 
S and D electrodes. \mbox{\boldmath$\epsilon_n$}$=\mbox{diag}(\epsilon_{n\uparrow},\epsilon_{n\downarrow})$ and
\mbox{\boldmath$t$}$=\mbox{diag}(t,t)$, where $t$ is the NNH strength, and $\epsilon_{n\sigma}$ represents the site energy
of an electron in the ring in absence of any magnetic interaction at site $n$ with spin $\sigma$ ($=\uparrow,\downarrow$).
\mbox{\boldmath$c_n$} contains the usual fermionic operators $c_{n\sigma}$. The most important term of the above system Hamiltonian
is \mbox{\boldmath $\vec{h}_n.\vec{\sigma}$} which is responsible for the spin dependent scattering. \mbox{\boldmath$\sigma$} 
is the vector containing Pauli spin matrices (assuming \mbox{\boldmath$\sigma_z$} as diagonal), and $h_n$ (=$|\mbox{\boldmath$\vec{h}_n$}|$) 
represents the spin-flip parameter. The orientation of \mbox{\boldmath$\vec{h}_n$} can in general be described by the usual spherical 
polar co-ordinates $\theta_n$ and $\varphi_n$. The strength of $h_n$ is usually very large compared to all other 
spin-dependent scattering
factors, which thus yields a large separation between the up and down spin channels~\cite{exch1}. This is one of the important motivations 
behind the consideration of a magnetic ring instead of a SO coupled one, like previous contemporary works. For comprehensive analysis 
of this of kind spin dependent scattering, see Refs.~\cite{exch1,exch2}. Finally, the Hamiltonian $H_T$ that represents the coupling 
between the ring and contact electrodes is given by
\begin{equation}
H_T = \left(\mbox{\boldmath$c_p^{\dagger} t_{S(D)} c_q$} + h.c. \right)
\label{eq4}
\end{equation}
where \mbox{\boldmath$t_{S(D)}$}$=\mbox{diag}(t_{S(D)},t_{S(D)})$. $t_S$ and $t_D$ denote the coupling strengths of the ring with 
S and D, respectively. $p$ and $q$ denotes the lattice points (those are variable) where S and D are attached to the ring.

This is all about the system Hamiltonian. Using this TB Hamiltonian (Eq.~\ref{eq1}) we need to find the required quantities to get the
logical operations. The key quantities include (i) bias driven circular current and (ii) induced magnetic field due to this current.
We evaluate the first one by determining bond current densities using the wave-guide (WG) theory within the TB framework. In this
methodology, a set of coupled equations involving wave amplitudes at different lattice sites of the ring along with the end sites 
of the electrodes with which the ring is directly coupled are solved. 
The equations are governed from the Schr\"{o}dinger equation $H|\phi\rangle = E|\phi\rangle$, where 
$|\phi\rangle = \sum_i C_{i\sigma} c_{i\sigma}^{\dagger}|0\rangle$. $C_{i\sigma}$'s are the wave amplitudes. The detailed description
of the equations is available in Appendix.~\ref{aa}, where the set of coupled equations are written based on the mathematical prescription
given in our recent work~\cite{cite14}. Here it is relevant to note that, the generalized wave-guide theory in the `spin basis' 
has been described for the first time by us in this work (Ref.~\cite{cite14}).
	
Measuring the bond current density following the steps given in Appendix~\ref{aa}, we determine current by integrating 
the current density using the relation
\begin{equation}
I_{n,n+1}(V) = \int\limits_{-\infty}^{\infty}J_{n,n+1}(E) [f_S(E) - f_D(E)]\, dE
\label{eq6}
\end{equation}
where \[f_{S(D)}=\frac{1}{1 + e^{(E-\mu_{S(D)})/{k_B \mathcal{T}}}}\] is the Fermi function for S(D), $k_B$ is the Boltzmann constant 
and $\mathcal{T}$ is the equilibrium temperature. $\mu_S$ and $\mu_D$ are the electro-chemical potentials and they are expressed as
$\mu_S=E_F + eV/2$ and $\mu_D=E_F - eV/2$, respectively, where $V$ is the bias voltage and $E_F$ is the equilibrium Fermi energy.

Using the individual bond currents, the net circular current is obtained through the sum~\cite{cite10,cite14}
\begin{equation}
I_C=\frac{1}{N} \sum_n I_{n,n+1}
\label{eq7}
\end{equation}
where $N$ denotes the total number of lattice sites in the ring. In determining the current we assign a positive sign for 
the current propagating in the anti-clockwise direction.

The above expression i.e., Eq.~\ref{eq7} is one of the required quantities to analyze the logical operations, as mentioned. Now, the
other key factor that needs to be evaluated is the induced magnetic field due to the circular current generated in the loop. The
magnetic field at any arbitrary point (say) $r$ is obtained using the well known Biot-Savart's law~\cite{cite10,cite14}
\begin{equation}
\vec{B}(\vec{r}) = \sum_n \left(\frac{\mu_0}{4\pi}\right)
\int I_{n,n+1}\frac{d\vec{r^{\prime}} \times(\vec{r}-\vec{r^{\prime}})}
{|(\vec{r}-\vec{r^{\prime}})|^3}
\label{eq8}
\end{equation}
where $\mu_0$ is the magnetic constant.

In our analysis, the logical outputs (ON and OFF states) are described by the orientation of the local (free) spin placed at 
the ring center. The spin is initially aligned along a particular direction which we refer as `OFF' state or low output, while in 
the presence of induced magnetic field when the free spin is aligned along $+Z$ direction we get the `ON' state or high output. We 
calculate the angle of rotation $\theta_C$ because of the magnetic field $B$, assuming the operation time $\tau$, from the 
condition~\cite{cite22,skm1}
\begin{equation}
\theta_C=\frac{g \mu_{\mbox{\tiny B}} B \tau}{2\hbar}
\label{eq9}
\end{equation}
where $g$ is the Lande $g$-factor and $\mu_{\mbox{\tiny B}}$ is the Bohr magneton. Thus, if the relative angle $\theta_C$ is zero, the 
logical output becomes `low', whereas the output is `high' when the spin is aligned along $Z$ direction.

Finally, to estimate the performance of logic circuits we compute power, delay and the product of these two. The 
power is easily determined, while the delay time is measured from the relation~\cite{delay1}
\begin{equation}
\tau_d=\frac{\left(Q_{\mbox{on}}-Q_{\mbox{off}}\right)}{I_{\mbox{on}}}
\label{del}
\end{equation}
where $Q_{\mbox{on}}$ and $Q_{\mbox{off}}$ represent the charges in the ON and OFF states, respectively, and $I_{\mbox{on}}$ corresponds
to on-current.

The charge $Q_{\mbox{on(off)}}$ at the bias voltage $V$ is obtained from the expression~\cite{delay2}
\begin{equation}
Q_{\mbox{on(off)}}=C_{\mbox{on(off)}} V = |e|^2 \rho_{\mbox{on(off)}} V
\label{chrg}
\end{equation}
where $C_{\mbox{on(off)}}$ is the quantum capacitance and $\rho_{\mbox{on(off)}}$ is the density of states.

\section{Numerical Results and Discussion}

The results are essentially arranged in two parts. One includes the individual logical operations, those are obtained by reconfiguring
a single setup. The other part describes the parallel logic options. Both are very important and here we discuss them one by one.
Before analyzing the results, let us mention the common parameter values those are taken into consideration through out the numerical 
calculations. We consider a six-site ring as a basic building block. In the ring, the on-site energies 
$\epsilon_{n\uparrow}=\epsilon_{n\downarrow}=0\,\forall \, n$, and the NNH hopping strength $t=0.5\,$eV. These quantities 
($\epsilon_0$ and $t_0$) for the side attached electrodes are taken as $0$ and $1\,$eV, respectively. The ring-to-electrode coupling 
strengths $t_S=t_D=0.25\,$eV. The spin flip parameter $h_n$ for all the sites is also fixed at $0.25\,$eV. All the 
physical pictures of
logic operations presented here remain unchanged for other choices of the parameter values as well, which we confirm through our rigorous 
numerical calculations. We take $\varphi_n=0$ for all $n$. Unless stated otherwise, we compute all the results at absolute zero 
temperature and at $E_F=0$.

As the central operations are associated with the orientation of the free spin placed at the ring center due to the 
induced magnetic
\begin{figure}[ht]
{\centering\resizebox*{8cm}{5.25cm}{\includegraphics{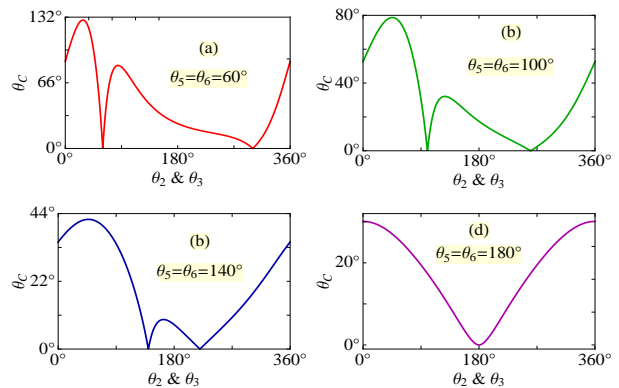}}\par}
\caption{(Color online). Angle of the rotation $\theta_C$ (measured in degree) of the free spin as a function of $\theta_2$ ($=\theta_3$) 
for four different cases of $\theta_5$ ($=\theta_6$). The bias voltage is set at $V=0.25\,$V.}
\label{fig2}
\end{figure}
\begin{figure*}[ht]
{\centering\resizebox*{16cm}{11cm}{\includegraphics{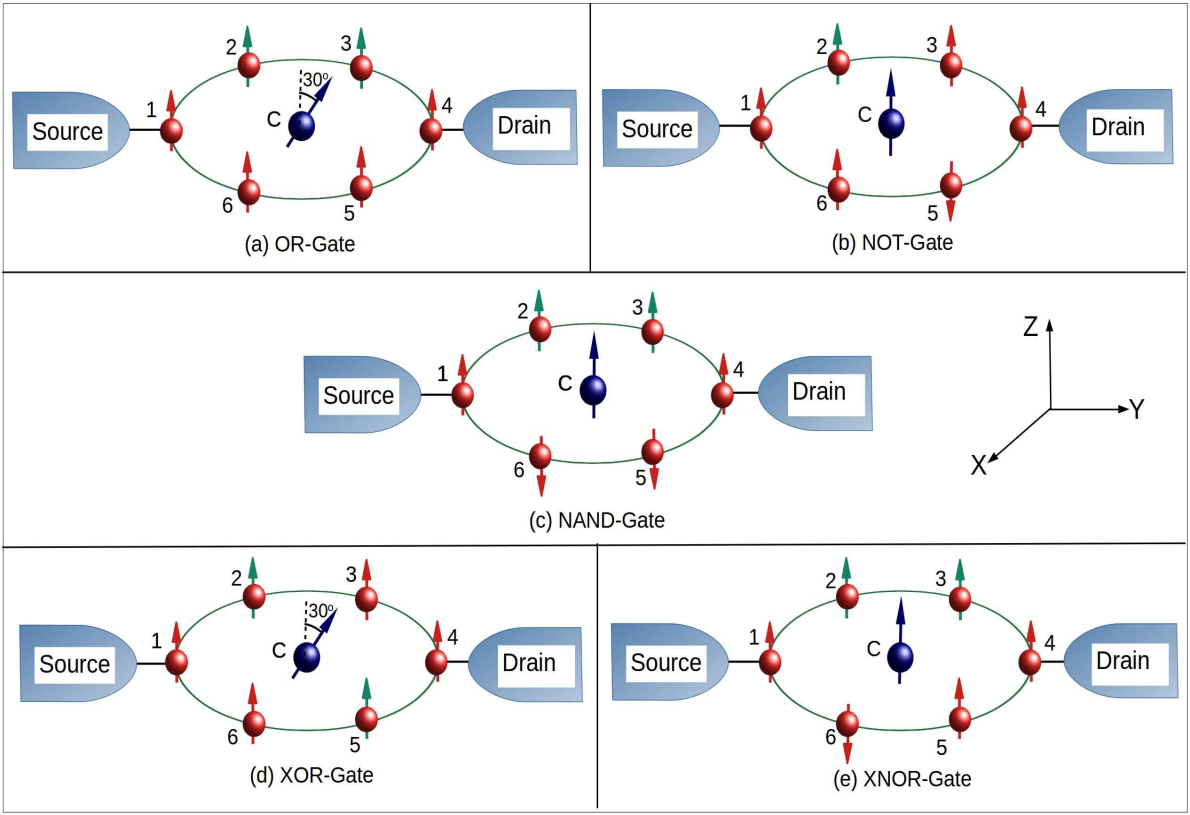}}\par}
\caption{(Color Online). Layouts of five different logical operations. The input conditions are imposed through the spins 
shown by the
green color. When the spin is up it indicates $0$ input state, while for the down spin the input state becomes $1$. The free spin is
initially orientated at $30^{\circ}$ which describes the zero output condition. All these logical operations are in principle designed
by reconfiguring a common basic structure.}
\label{fig3}
\end{figure*}
field, first it is important to check the response of this free spin viz, the possible degree of rotation under different 
input conditions.
More specifically, we want to examine whether the induced magnetic field is suitable enough to rotate the spin or not. The results 
are shown in Fig.~\ref{fig2} considering a similar kind of setup as taken in Fig.~\ref{fig1}. We fix the alignments of spins at sites 5
and 6 (measured by $\theta_5$ and $\theta_6$), and rotate the spins at 2 and 3 (measured by $\theta_2$ and $\theta_3$) keeping them
in parallel i.e., $\theta_2=\theta_3$. Here we choose $\theta_1=\theta_4=0$, and set the bias $V=0.25\,$V. A finite circular current,
and thus magnetic field, is generated as long as the net current passing through one arm is different from the current flowing in the
other arm, since the arm lengths are equal for our nano junction. 
Four different cases are shown, depending on the typical values of $\theta_5$ ($=\theta_6$). In each of these 
cases the angle of rotation $\theta_C$ of the free spin is determined, by measuring the magnetic field associated with $I_C$, 
following the relation given in Eq.~\ref{eq9}. The orientations of the spins located at sites 2 and 3 are varied from $0$ to $2\pi$.
Interestingly we see that the free spin can have a high degree of rotation (in one case it becomes more than $130^{\circ}$),
depending on $\theta_5$ ($=\theta_6$). This large degree of rotation is possible due to the generation of much higher magnetic field
at the ring center. It clearly suggests that, the required magnetic field to rotate a free spin can easily be achieved, and this higher 
magnetic field is obtained mainly because of the smaller loop area of the ring. Thus, we can safely utilize this prescription to 
recognize two different states of the output. The vanishing rotation ($\theta_C=0$) is observed only for the two typical situations. 
These are (i) $\theta_2$ ($=\theta_3$) = $\theta_5$ ($=\theta_6$) and (ii) $\theta_2$ ($=\theta_3$) = $2\pi-\theta_5$ ($=2\pi-\theta_6$).
Under these conditions, the currents through the upper and lower arms of the ring become identical, resulting vanishing $I_C$ and
hence the induced magnetic field $B$.
\begin{figure*}[ht]
{\centering\resizebox*{12cm}{10cm}{\includegraphics{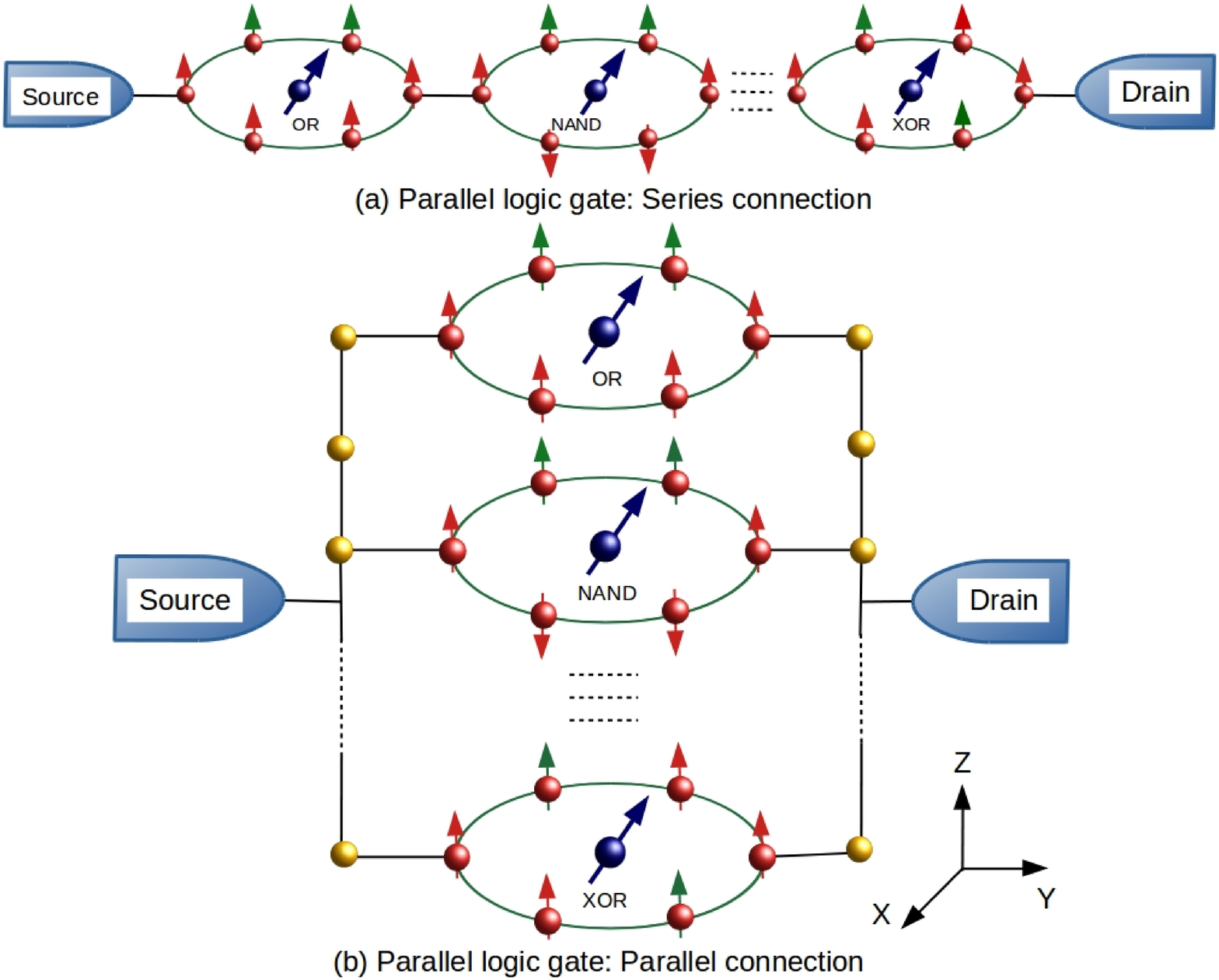}}\par}
\caption{(Color Online). Outline to achieve all possible logical operations in a single nanojunction device, where individual loops 
are used to serve different logic gates. All the colored spins correspond to the identical meaning as discussed in Fig.~\ref{fig3}.
The parallel operations are performed in `two ways' by connecting the loops in (a) series and (b) parallel. In each of these diagrams,
the dotted portion implies that we can add any number of rings representing different and/or identical logical operations in series 
or parallel.}
\label{fig4a}
\end{figure*}
\begin{figure*}[ht]
{\centering\resizebox*{15cm}{11cm}{\includegraphics{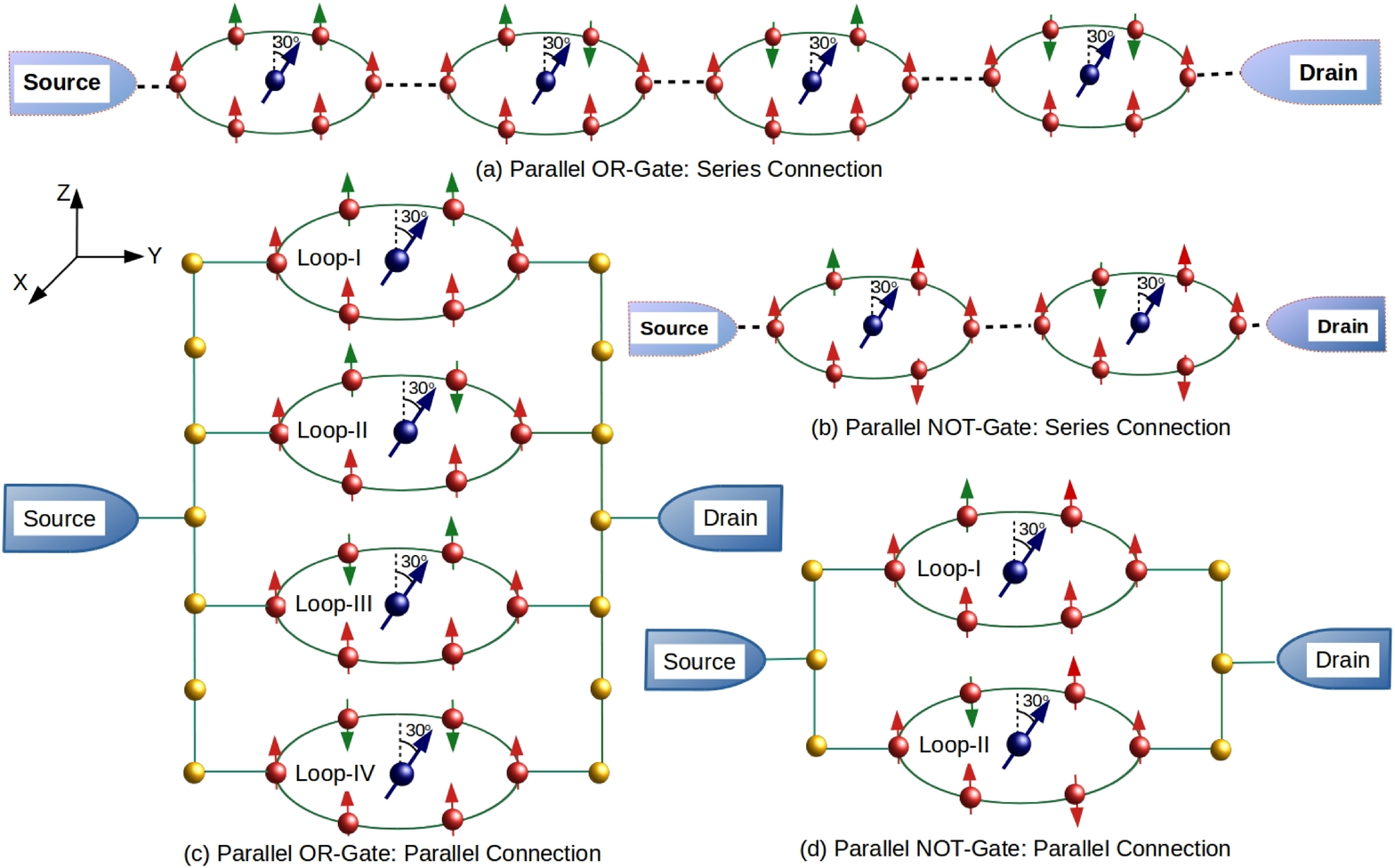}}\par}
\caption{(Color Online). Different mechanisms of getting parallel operations. All the inputs of a logic gate are implemented at one
time, and all the outputs are read simultaneously. For a two-input logic gate total four loops, whereas for the two-input gate two
loops are considered depending on the possible input conditions. We arrange them either in series or parallel, as schematically shown 
here. The different colored spins carry the usual meanings as mentioned earlier in other figures. Among all the two-input logic gates, 
here we show the operation of OR gate, as a typical example. Similar prescription is followed for the other two-input gates.}
\label{fig4}
\end{figure*}

\subsection{Reprogrammable spin logic gate}

With the above verification that a sufficient magnetic field can be generated at the ring center to rotate a free spin, 
now we explain the logical operations. Figure~\ref{fig3} shows the setups of designing five logic gates, OR, NOT, NAND, XOR and XNOR,
where all these logic gates are re-configured from a common basic structure. The spins shown by the green color are used to
regulate the input states. When the spin is up we get $0$ input state, while for down spin the input state becomes $1$, and we
follow this prescription throughout the discussion. The free spin, amended at the ring center, is aligned at an angle $30^{\circ}$
with respect to the $Z$ axis which represents the `low' output state, and when it gets oriented along $Z$ direction in presence of 
finite magnetic field at the ring center we get `high' output state. Now, in order to rotate the spin by $30^{\circ}$,
the minimum magnetic field that is required is $\sim 2.4\,$mT (calculated by using Eq.~\ref{eq9}).
We measure the response i.e., the logical outputs in terms of the magnetic field under different input conditions. The results are
described in a tabular form (see Table~\ref{tab11}) to have a clear picture of all the logical operations.

\begin{table}[ht]
\caption{Truth tables for different logical operations. Output is described in terms of the generated magnetic field (measured in
unit of mT) at the loop center. The bias is fixed at $V=0.25\,$V.}
$~$
\vskip -0.25cm
\fontsize{7}{9}
\begin{tabular}{|c|c|c|c|c|c||c|c|}
\hline
\multicolumn{2}{|c|}{Input} &
\multicolumn{4}{c||}{Output} &
\multicolumn{2}{c|}{NOT}\\
\cline{3-8}
\multicolumn{2}{|c|}{} & OR & NAND & XOR & XNOR & Input & Output\\
 \hline
 $\uparrow$ & $\uparrow$ & 0 & 2.5 & 0 & 40 & $\uparrow$ & 40 \\
 $\uparrow$ & $\downarrow$ & 40 & 2.5 & 40 & 0 &$\downarrow$ & 0 \\
 $\downarrow$ &$\uparrow$ & 40 & 2.5 & 40 & 0 & & \\
 $\downarrow$ &$\downarrow$ & 2.5 & 0 & 0 & 2.5 & & \\
 \hline
\end{tabular}
\label{tab11}
\end{table}
The selection of the input spin(s) is very crucial, as we need to satisfy the required `symmetric' and `asymmetric' 
conditions for
getting zero and non-zero circular currents, and thus, induced magnetic field, to have all these kinds of logical operations.
For instance, if we focus on the OR gate setup, we see that the spins at sites 2 and 3 are used to set the logical inputs, whereas
for XOR gate the spins at sites 2 and 5 are considered. Not only that, we need to concentrate on the other spins as well (drawn by the
red color) located at other sites, to satisfy the required conditions. But, the fact is that we develop all these logical
operations by re-configuring a single setup. This is quite interesting and important as well which we strongly believe, and may
give some impact in designing future nanoelectronic devices.

\subsection{Parallel logic operations: Two prescriptions}

The above sub-section (Sub-Sec. B) illustrates single logic operation at a particular time. As all the logic functions are based on the 
operation of the free spin due to the induced magnetic field at the ring center, we can easily extrapolate this idea to implement several
logic operations simultaneously, as many as we wish, in a single setup. Under this case, the performance of one logic gate is no longer 
disturbed by the others, which is the most important issue of our analysis. This feature is no longer expected from 
the earlier propositions of logic operations. Simultaneous operations can be obtained but they are highly limited, as in most of the cases
one logic function can be influenced by the other. All these issues are successfully eliminated in our present proposal. Two separate 
schemes are explored to execute the parallel operations those are: (i) all logic gates are operated simultaneously in one setup and (ii) 
all possible inputs are given at a particular time and all the output signals are read simultaneously. These aspects are analyzed one by 
one as given below.

\subsubsection{Designing of different logic gates in one setup}

Here we emphasize on how to perform different logical operations at an identical time in a single junction device. It can be done
in two ways, those are schematically illustrated in Figs.~\ref{fig4a}(a) and (b), where the loops, involving different logic gates,
are connected in series and parallel, respectively. For individual logic gates, the architecture should be followed as discussed in
Fig.~\ref{fig3} which means we need to satisfy the identical `symmetric' and `asymmetric' conditions to get the operations. For the 
parallel connection of the ring conductors, it may seem that the circuit diagram is quite trivial and expected as well, since all these
individual loops are getting identical power from the contact electrodes, like what we have for different functional elements in 
conventional CPUs in PCs. But, for better clarity and clear understanding we expose both the two possibilities (series and parallel) for
simultaneous logic operations. Unlike parallel connection, the loops are not getting identical power when they are coupled in series, 
although we can fully satisfy our purpose. Thus, designing of all possible logic gates at a same time is of course possible.

\subsubsection{Parallel operations in a single logic gate}

Designing of a logic gate where all the inputs are given in parallel and all the outputs are also read simultaneously is undoubtedly 
extremely significant. So far only one work is available~\cite{cite16a} along this line, though the functional operations described in 
that work are not spin based. Now, to satisfy the parallelized conditions in our spin based device, we have to couple four or two loops 
depending on whether it is two-input or one-input logic gate, as for a two-input gate there are four possible input (output) conditions,
\begin{table}[ht]
\caption{Truth tables for different parallel logical operations where the loops are connected in series configuration. Output response 
is described in terms of the induced magnetic field (measured in unit of mT) at the loop centers.}
$~$
\vskip -0.25cm
\fontsize{7}{9}
\begin{tabular}{|c|c|c|c|c|c|c||c|c|}
\hline
& \multicolumn{2}{c|}{Input} &
\multicolumn{4}{c||}{Output} &
\multicolumn{2}{c|}{NOT}\\
\cline{4-9}
& \multicolumn{2}{c|}{} & OR & NAND & XOR & XNOR & Input & Output\\
 \hline
 Loop-1 & $\uparrow$ & $\uparrow$ & 0 & 49 & 0 & 94 & $\uparrow$ & 4.7\\
 Loop-2 & $\uparrow$ & $\downarrow$ & 4.4 & 14 & 10 & 0 &$\downarrow$ & 0 \\
 Loop-3 & $\downarrow$ &$\uparrow$ & 4.4 & 14 & 10 & 0 & & \\
 Loop-4 & $\downarrow$ &$\downarrow$ & 4.9 & 0 & 0 & 8.4 & & \\
 \hline
\end{tabular}
\label{tab1}
\end{table}
while for the other case two input (output) conditions are there. Among all the possible two-input gates, here we take 
OR gate as a
prototype example to explain the parallel operation, and all the other such gates can be described in the same footing. Together with
OR gate, we also
\begin{table}[ht]
\caption{Same as Table~\ref{tab1}, when the loops are arranged in parallel configuration.}
$~$
\vskip -0.25cm
\fontsize{7}{9}
\begin{tabular}{|c|c|c|c|c|c|c||c|c|}
\hline
& \multicolumn{2}{c|}{Input} &
\multicolumn{4}{c||}{Output} &
\multicolumn{2}{c|}{NOT}\\
\cline{4-9}
& \multicolumn{2}{c|}{} & OR & NAND & XOR & XNOR & Input & Output\\
 \hline
 Loop-1 & $\uparrow$ & $\uparrow$ & 0 & 8.9 & 0 & 5 & $\uparrow$ & 16.2\\
 Loop-2 & $\uparrow$ & $\downarrow$ & 18.5 & 9.4 & 13.8 & 0 &$\downarrow$ & 0 \\
 Loop-3 & $\downarrow$ &$\uparrow$ & 14 & 5.8 & 20 & 0 & & \\
 Loop-4 & $\downarrow$ &$\downarrow$ & 3.6 & 0 & 0 & 6.8 & & \\
 \hline
\end{tabular}
\label{tab2}
\end{table}
consider NOT gate to discuss the operation for the one-input logic gate.

The schematic arrangements of the loops for designing the parallel OR and NOT gates are shown in Fig.~\ref{fig4}. Both the series 
and parallel configurations of the loops are taken into account, where each loop is associated with one logical input condition, and 
the output is measured by the response of the free spin embedded at the ring center, like previous operations. Both for the series 
and parallel configurations, the response of any loop gets no longer disturbed by the responses of the other loops. They work well 
independently, which is the notable feature of our prescription. These logical functions cannot be performed if the operations are 
made using the concept of the conventional transport current, since the vanishing transport current across any loop stops the current
\begin{figure}[ht]
{\centering\resizebox*{8cm}{7cm}{\includegraphics{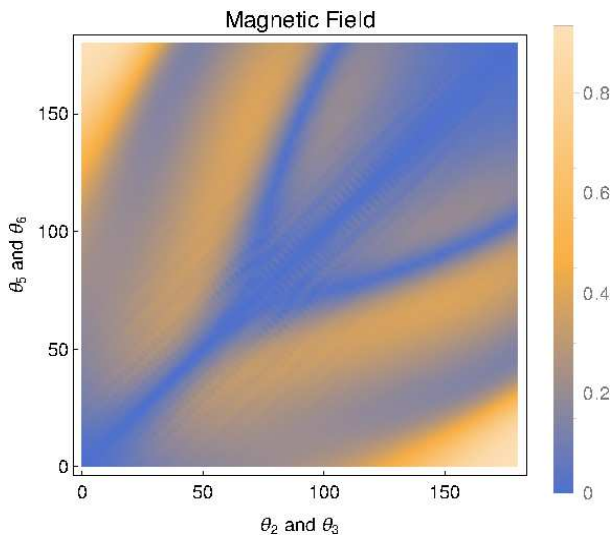}}\par}
\caption{(Color Online). Density plot of induced magnetic field at the ring center, associated with the circular current, as functions
of $\theta_2$ ($=\theta_3$) and $\theta_5$ ($=\theta_6$). Here we take the nanojunction as shown in Fig.~\ref{fig1}, and compute the
results at the bias voltage $V=2\,$V. The system temperature is fixed at $\mathcal{T}=300\,$K.}
\label{fig5}
\end{figure}
flow through all other loops, which essentially gives a null state (i.e., the circuit becomes off). On the other hand, 
the vanishing 
circular current in any loop under the symmetric condition does not yield vanishing transport current, and therefore, we can safely
perform the required operations.

The responses of different logic gates are shown in tabular form. In Table~\ref{tab1}, the results are presented when the loops are 
configured in series. To have better output, we fix the bias voltage $V=0.25\,$V for the OR, XOR and NOT gates, while for the other
two gates (NAND and XNOR) we set $V=0.7\,$V. These biases are not specific, and any other voltage can also be taken into account. 
For the case of parallel configuration of the rings, the results are given in Table~\ref{tab2}. Here we choose the biases as $V=0.6\,$V 
and $0.7\,$V like the above sets of logic gates. Both from Table~\ref{tab1} and Table~\ref{tab2}, the logical responses are clearly
reflected, and thus we can emphasize that the parallel operations can be substantiated where all the inputs are switched on together
and outputs are read at an identical time.

\subsection{Effect of temperature}

All the results presented above are worked out at zero temperature. Keeping in mind the realistic situation, now we 
want to examine 
whether the above facts are still valid even at moderate temperature or not. This checking is always crucial for device designing.
The another fact is that, in the above analysis we consider up and down spin orientations as the two input states, and thus 
we align other spins to get the required symmetric and asymmetric conditions to establish the logical operations. Instead of these, 
one can also consider any other arrangements of the input states and accordingly we need to align the other spins to reach the 
required conditions. In any such arrangements we have to make sure that the induced magnetic field is sufficient enough to rotate 
the free spin in desired angle to achieve the high output state.

In Fig.~\ref{fig5}, we show a density plot of the induced magnetic field at the ring center, considering a similar 
kind of nanojunction  
as taken in Fig.~\ref{fig1}, by simultaneously varying $\theta_2$ ($=\theta_3$) and $\theta_5$ ($=\theta_6$). The results are computed 
at the bias voltage $V=2\,$V, setting the temperature $\mathcal{T}=300\,$K. Clearly we see that, for a wide range of $\theta_2$ and
$\theta_5$ considerably large magnetic field is induced because of the non-zero circular current in the ring. At finite temperature,
since all the energy levels contribute to the current, there is a possibility to get smaller spin dependent circular current due to
mutual cancellations, and hence the magnetic field. But for our case we still get a reasonably large circular current to satisfy the 
required logic operations. Thus, all the logical operations stated here can be achieved in realistic situation and we believe that 
the proposal can be verified in suitable laboratories.

\subsection{Power, delay and power-delay-product}

Consumption of lower power and much faster operation i.e., lesser delay time are the two important prerequisites 
to get an efficient device. The efficiency can be quantified in a more general way by calculating the product of the power and the 
delay time. The PDP is called as the {\em figure of merit} which measures the energy consumed in each operation.
\begin{table}[ht]
\caption{Power, delay, and PDP of various logic devices at $V=0.25\,$V and $E_F = 0.75\,$eV.}
$~$
\vskip -0.25cm
\fontsize{7}{9}
\begin{tabular}{|c|c|c|c|}
\hline
 & Delay (s) & Power (W) & PDP (J) \\
\hline
OR Gate & $7.78\times10^{-16}$ & $1.55\times10^{-8}$ & $1.21\times10^{-23}$\\
Parallel OR Gate I & $2.24\times10^{-12}$ & $9.07\times10^{-14}$ & $2.03\times10^{-25}$\\
Parallel OR Gate II & $3.86\times10^{-15}$ & $5.28\times10^{-11}$ & $2.04\times10^{-25}$\\
\hline
\end{tabular}
\label{tab3}
\end{table}
What we find from our extensive calculations is that the PDP is extremely small for all the logical operations studied 
in this work. In Table~\ref{tab3} we present the results of some of them, and they are single OR gate, parallel OR gate for the series 
connection (which we refer in the table as parallel OR gate I), and parallel OR gate for the parallel connection (which is mentioned in 
the table as parallel OR gate II). For other logic gates the delay, power and PDP are of the same orders of magnitude as given in 
Table~\ref{tab3}.
All the quantities are very low compared to the traditional TTL and CMOS logic families. The underlying mechanism relies on the fact that all
the operations discussed here are performed based on the `spin states' unlike the conventional logic functions where charges are taken into 
account for the logic inputs and/or outputs. The factors shown in Table~\ref{tab3} remain almost unchanged even for much higher bias voltages
which we confirm through our numerous numerical calculations. With these results we can strongly argue that the proposed logic operations are
superior than the existing logic families.

\subsection{Experimental realization of magnetic nano rings}

Finally, we discuss how to realize magnetic nano rings experimentally those are used to have logical operations. 
Several prescriptions
are available in literature and with the help of those advanced methodologies it is now possible to get simple and complex patterned 
magnetic ring geometries. For instance here we want to highlight some of those approaches for the benefit of readers. The most common
techniques to design magnetic rings are the electron beam lithography~\cite{ebl}, silicon etching processes~\cite{sep} and the 
nanosphere lithography~\cite{nsl1,nsl2}. Sometimes droplet epitaxy method~\cite{dem} is also used for the fabrication of ring like 
geometries. Using this prescription, Mano {\em et al.}~\cite{mano} have designed magnetic rings with diameters of the order of few
nm, and also suggested how to tune the size of the magnetic rings. Yang and co-workers have claimed in their work that more controlled
magnetic rings can be constructed by dewetting magnetic nanoparticle solution in a specific substrate~\cite{dew}. With these enormous
possibilities, we hope that magnetic nano rings of desired radius can be fabricated, and the orientations of selective spins at distinct
magnetic sites can be controlled selectively following the prescriptions discussed earlier. Thus our propositions of logic operations using
magnetic rings can be tested in a suitable laboratory setup.

\section{Closing Remarks}

The work essentially deals with the possible designing mechanisms of logic gates at nano-scale level, where everything is `spin based',
circumventing the traditional prescription of getting `charge based' ones. Few contemporary proposals are already available, though such 
a comprehensive analysis, presented here, using the concept of `bias driven spin based circular current in a nano loop' does not exist
in literature to the best of our concern. They key aspects and finding of our work are summarized as follows.\\
$\bullet$ Before presenting the results, we have checked whether the circular current induced magnetic field is sufficient enough 
to rotate a free spin in desired angle. What we have found is that, a high degree of rotation can be made as the induced magnetic field
is considerably large. This high magnetic field is achieved mainly due to the small size ring. \\
$\bullet$ Following this verification, we have discussed how to substantiate logical operations. Total five logic gates have been 
designed. \\
$\bullet$ In the rest part of our analysis we have mostly focussed on the principle of getting parallel logic operations. Here two 
different prescriptions have been proposed. In one case all the logic gates can be operated simultaneously, while in the other 
prescription a single logic gate has been taken into account where all the inputs can be implemented at a same time and also all the 
outputs can be read simultaneously. Both these two proposals are interesting and important as well.\\
$\bullet$ In the construction of parallel operations, the most notable point is that one logical operation gets no longer disturbed by 
the other. Since the operations depend entirely on the generated circular current, not on the transport current, we can get simultaneous
operations from individual nano rings, whether they are connected in series or parallel. \\
$\bullet$ We have also discussed the effects of temperature, and found that all the logical operations are equally valid even at moderate 
temperatures. \\
$\bullet$ We have tested the efficiency of the logic circuits by studying power, delay and PDP. From our results we have
reached to the conclusion that all the logic functions studied here are quite superior than the conventional logic families. \\
$\bullet$ Finally, we have given a brief outline about the designing principle of nano rings, for the sake of 
completeness of our analysis.

We end by stating that, the present analysis can give a boost in the field of reconfigurable logic operations since all the logic
gates are designed by rearranging the spins in a single setup. The device in one one hand is very simple to understand and on the other 
hand it can probably be designed quite easily in advanced laboratories.

\section*{ACKNOWLEDGMENT}

SKM would like to thank the financial support of DST-SERB, Government of India, through the Project Grant Number EMR/2017/000504. 

\appendix

\section{Calculation of current density from the set of coupled equations}
\label{aa}

Both for incident up and down spin electrons, we need to write a separate set of coupled equations. Let us begin with 
an up spin electron which gets injected from the source end to the ring system as a plane wave with unit amplitude. For the nanojunction 
shown Fig.~\ref{fig1}, the set of coupled equations are expressed, in general forms
(considering $\epsilon_{n\uparrow}=\epsilon_{n\downarrow}=\epsilon_{n}$) as follows:
{
\allowdisplaybreaks
\begin{widetext}
{\footnotesize
\begin{eqnarray}
\left[\left(\begin{array}{cc}
        E & 0 \\
        0 & E
\end{array}\right) - \left(\begin{array}{cc}
    \epsilon_0 & 0 \\
    0 & \epsilon_0
\end{array}\right)\right]\left(\begin{array}{cc}
        1 + \rho_{\uparrow\uparrow} \\
        \rho_{\uparrow\downarrow}
\end{array}\right)  = \left(\begin{array}{cc}
    t_0 & 0 \\
    0 & t_0
\end{array}\right) \left(\begin{array}{cc}
    e^{-ika} + \rho_{\uparrow\uparrow}e^{ika} \\
        \rho_{\uparrow\downarrow}e^{ika}
    \end{array}\right) + \left(\begin{array}{cc}
   t_{S}  & 0 \\
    0 & t_{S}
\end{array}\right) \left(\begin{array}{cc}
   C_{1,\uparrow\uparrow}  & 0 \\
    0 & C_{1,\uparrow\downarrow}
\end{array}\right)\nonumber \\
\left[\left(\begin{array}{cc}
    E & 0 \\
    0 & E
\end{array}\right) - \left(\begin{array}{cc}
    \epsilon_1 + h_1\cos\theta_1 & \sin\theta_1e^{-i\varphi_1} \\ 
    \sin\theta_1e^{i\varphi_1} & \epsilon_1 - h_1\cos\theta_1
\end{array}\right)\right] \left(\begin{array}{cc}
   C_{1,\uparrow\uparrow}  & 0 \\
    0 & C_{1,\uparrow\downarrow}
\end{array}\right) = \left(\begin{array}{cc}
    t_{S} & 0 \\
    0 & t_{S}
\end{array}\right)\left(\begin{array}{cc}
    1 + \rho_{\uparrow\uparrow(S)} \\
    \rho_{\uparrow\downarrow(S)}
    \end{array}\right) + \left(\begin{array}{cc}
    t & 0 \\
    0 & t
\end{array}\right) \left(\begin{array}{cc}
   C_{2,\uparrow\uparrow}  & 0 \\
    0 & C_{2,\uparrow\downarrow}
\end{array}\right) \nonumber\\
+ \left(\begin{array}{cc} 
 t & 0 \\
    0 & t
\end{array}\right) \left(\begin{array}{cc}
   C_{6,\uparrow\uparrow}  & 0 \\
    0 & C_{6,\uparrow\downarrow}
\end{array}\right)
\nonumber \\
\left[\left(\begin{array}{cc}
    E & 0 \\
    0 & E
\end{array}\right) - \left(\begin{array}{cc}
    \epsilon_2 + h_2\cos\theta_2 & \sin\theta_2e^{-i\varphi_2} \\ 
    \sin\theta_2e^{i\varphi_2} & \epsilon_2 - h_2\cos\theta_2
\end{array}\right)\right] \left(\begin{array}{cc}
   C_{2,\uparrow\uparrow}  & 0 \\
    0 & C_{2,\uparrow\downarrow}
\end{array}\right) = \left(\begin{array}{cc}
    t & 0 \\
    0 & t
\end{array}\right) \left(\begin{array}{cc}
   C_{1,\uparrow\uparrow}  & 0 \\
    0 & C_{1,\uparrow\downarrow}
\end{array}\right) + \left(\begin{array}{cc}
    t & 0 \\
    0 & t
\end{array}\right) \left(\begin{array}{cc}
   C_{3,\uparrow\uparrow}  & 0 \\
    0 & C_{3,\uparrow\downarrow}
\end{array}\right)\nonumber \\
\left[\left(\begin{array}{cc}    
    E & 0 \\
    0 & E
\end{array}\right) - \left(\begin{array}{cc}
    \epsilon_3 + h_3\cos\theta_3 & \sin\theta_3e^{-i\varphi_3} \\ 
    \sin\theta_3e^{i\varphi_3} & \epsilon_3 - h_3\cos\theta_3
\end{array}\right)\right] \left(\begin{array}{cc}
   C_{3,\uparrow\uparrow}  & 0 \\
    0 & C_{3,\uparrow\downarrow}
\end{array}\right) = \left(\begin{array}{cc}
    t & 0 \\
    0 & t
\end{array}\right) \left(\begin{array}{cc}
   C_{2,\uparrow\uparrow}  & 0 \\
    0 & C_{2,\uparrow\downarrow}
\end{array}\right) + \left(\begin{array}{cc}
    t & 0 \\
    0 & t
\end{array}\right) \left(\begin{array}{cc}
   C_{4,\uparrow\uparrow}  & 0 \\
    0 & C_{4,\uparrow\downarrow}
\end{array}\right) \nonumber \\
\left[\left(\begin{array}{cc}
    E & 0 \\
    0 & E
\end{array}\right) - \left(\begin{array}{cc}
    \epsilon_4 + h_4\cos\theta_4 & \sin\theta_4e^{-i\varphi_4} \\ 
    \sin\theta_4e^{i\varphi_4} & \epsilon_4 - h_4\cos\theta_4
\end{array}\right)\right] \left(\begin{array}{cc}
   C_{4,\uparrow\uparrow}  & 0 \\
    0 & C_{4,\uparrow\downarrow}
\end{array}\right)
  = \left(\begin{array}{cc}
    t & 0 \\
    0 & t
\end{array}\right) \left(\begin{array}{cc}
   C_{3,\uparrow\uparrow}  & 0 \\
    0 & C_{3,\uparrow\downarrow}
\end{array}\right) +  \left(\begin{array}{cc}
    t & 0 \\
    0 & t
\end{array}\right) \left(\begin{array}{cc}
   C_{5,\uparrow\uparrow}  & 0 \\
    0 & C_{5,\uparrow\downarrow}
\end{array}\right) \nonumber \\
+ \left(\begin{array}{cc}
    t_D & 0 \\
    0 & t_D
\end{array}\right)\left(\begin{array}{cc}
    \tau_{\uparrow\uparrow}e^{ika} \\
    \tau_{\uparrow\downarrow}e^{ika}
    \end{array}\right)\nonumber \\
\left[\left(\begin{array}{cc}    
    E & 0 \\
    0 & E
\end{array}\right) - \left(\begin{array}{cc}
    \epsilon_5 + h_5\cos\theta_5 & \sin\theta_5e^{-i\varphi_5} \\ 
    \sin\theta_5e^{i\varphi_5} & \epsilon_5 - h_5\cos\theta_5
\end{array}\right)\right] \left(\begin{array}{cc}
    C_{5,\uparrow\uparrow}  & 0 \\
    0 & C_{5,\uparrow\downarrow}
\end{array}\right) = \left(\begin{array}{cc}
    t & 0 \\
    0 & t
\end{array}\right) \left(\begin{array}{cc}
   C_{4,\uparrow\uparrow}  & 0 \\
    0 & C_{4,\uparrow\downarrow}
\end{array}\right) + \left(\begin{array}{cc}
    t & 0 \\
    0 & t
\end{array}\right) \left(\begin{array}{cc}
  C_{6,\uparrow\uparrow}  & 0 \\
    0 & C_{6,\uparrow\downarrow}
\end{array}\right) \nonumber \\
\left[\left(\begin{array}{cc}
    E & 0 \\
    0 & E
\end{array}\right) - \left(\begin{array}{cc}
    \epsilon_6 + h_6\cos\theta_6 & \sin\theta_6e^{-i\varphi_6} \\ 
    \sin\theta_6e^{i\varphi_6} & \epsilon_6 - h_6\cos\theta_6
\end{array}\right)\right] \left(\begin{array}{cc}
   C_{6,\uparrow\uparrow}  & 0 \\
    0 & C_{6,\uparrow\downarrow}
\end{array}\right)
  = \left(\begin{array}{cc}
    t & 0 \\
    0 & t
\end{array}\right) \left(\begin{array}{cc}
   C_{5,\uparrow\uparrow}  & 0 \\
    0 & C_{5,\uparrow\downarrow}
\end{array}\right) +  \left(\begin{array}{cc}
    t & 0 \\
    0 & t
\end{array}\right) \left(\begin{array}{cc}
   C_{1,\uparrow\uparrow}  & 0 \\
    0 & C_{1,\uparrow\downarrow}
\end{array}\right)\nonumber \\
\left[\left(\begin{array}{cc}
        E & 0 \\
        0 & E
\end{array}\right) - \left(\begin{array}{cc}
    \epsilon_0 & 0 \\ 
    0 & \epsilon_0
\end{array}\right)\right]\left(\begin{array}{cc}
        1 + \tau_{\uparrow\uparrow} \\
        \tau_{\uparrow\downarrow}
    \end{array}\right) = \left(\begin{array}{cc}
    t & 0 \\
    0 & t
\end{array}\right) \left(\begin{array}{cc}
   C_{4,\uparrow\uparrow}  & 0 \\
   0 & C_{4,\uparrow\downarrow}
\end{array}\right) + \left(\begin{array}{cc}
    t_0 & 0 \\
    0 & t_0
\end{array}\right) \left(\begin{array}{cc}
    \tau_{\uparrow\uparrow}e^{2ika} \\
    \tau_{\uparrow\downarrow}e^{2ika}
    \end{array}\right) \nonumber \\
\label{eq3a}
\end{eqnarray}}
\end{widetext}
The factors $\rho_{\sigma\sigma^{\prime}}$ and $\tau_{\sigma\sigma^{\prime}}$ correspond to the spin dependent reflection and 
transmission amplitudes, respectively, $k$ is the wave vector and $a$ is the inter-atomic distance. 

Solving Eq.~\ref{eq3a}, we get all the required co-efficients, and, then the bond current density between any bond 
connecting the sites $n$ and ($n+1$) is obtained from the relation
\begin{equation}
J_{n,n+1,\sigma\sigma^{\prime}}(E)=\frac{(2e/\hbar)\mbox{Im}\left[t\,C_{n,\sigma\sigma^{\prime}}^*
C_{i+1,\sigma\sigma^{\prime}} \right]}{(2e/\hbar)(1/2)t_0\sin(ka)}.
\label{eq5}
\end{equation}
Here $\sigma$ is used for the incident spin, while $\sigma^{\prime}$ denotes the transmitting spin.

In the same footing we can write another set of coupled equations for the down spin electron, and determining the co-efficients
we compute $J_{n\rightarrow n+1\downarrow\downarrow}$ and $J_{n\rightarrow n+1\downarrow\uparrow}$. 

Once we have all these components involving circular bond current densities, we finally obtain the net bond current density 
$J_{n,n+1}$ by the relation 
\begin{eqnarray}
J_{n,n+1}(E) & = & J_{n,n+1,\uparrow\uparrow} + J_{n,n+1,\uparrow\downarrow} \nonumber \\
 & & + J_{n,n+1,\downarrow\downarrow} + J_{n, n+1,\downarrow\uparrow}.
\label{apeq6}
\end{eqnarray}

The above prescription can now easily be generalized for any junction setup.

\end{document}